\documentclass{acm_proc_article-sp}
\usepackage{graphicx}
\usepackage{amssymb}
\usepackage{amsmath}
\usepackage{subfigure}
\usepackage{bm}
\usepackage{url}

\newcommand{\be}{\begin{equation}}
\newcommand{\ee}{\end{equation}}

\begin{document}
\title{VCG Payments for Portfolio Allocations in Online Advertising}
%\titlerunning{VCG for Portfolio Allocations}

\numberofauthors{4} %  in this sample file, there are a *total*
\author{
% 1st. author
\alignauthor
James Li\\
      \affaddr{Yahoo Labs}\\
      \email{jamesyili@yahoo-inc.edu}
\alignauthor
Eric Bax\\
      \affaddr{Yahoo Labs}\\
       \email{baxhome@yahoo.com}
\alignauthor
Nilanjan Roy\\
      \affaddr{Singapore University of Technology and Design}\\
       \email{nilu8603@gmail.com}
\and
\alignauthor 
Andrea Leistra\\
     \affaddr{LivingSocial}\\
     \email{a\_leistra@yahoo.com}
}

\maketitle

\begin{abstract}
\noindent Some online advertising offers pay only when an ad elicits a response. Randomness and uncertainty about response rates make showing those ads a risky investment for online publishers. Like financial investors, publishers can use portfolio allocation over multiple advertising offers to pursue revenue while controlling risk. Allocations over multiple offers do not have a distinct winner and runner-up, so the usual second-price mechanism does not apply. This paper develops a pricing mechanism for portfolio allocations. The mechanism is efficient, truthful, and rewards offers that reduce risk.
\keywords{online advertising, auction mechanism, portfolio allocation, VCG, uncertainty}
\end{abstract}

%\textbf{JEL codes:} 4.004: D4 - Market Structure and Pricing, 4.008: D8 - Information and Uncertainty

%\clearpage

\section{Introduction}

An \textit{ad call} is the opportunity to display an ad to a user on a web page view. In online advertising auctions, publishers sell ad calls and advertisers buy them. (For background on online advertising auctions, refer to Varian \cite{varian09,varian07}, Edelman, Ostrovsky, and Schwarz \cite{edelman07}, and Lahaie and Pennock \cite{lahaie07}.) Some advertisers bid for ad calls directly, but others submit offers to pay their bid only if the ad elicits a response from the user, such as a click or a purchase. For those offers, the expected offer value is the per-response bid times the response rate. So the auctioneer must estimate response rates to compare offers on the basis of expected value.

When advertisers pay based on responses, their offers have random payoffs. Since there are errors in estimating response rates, the payoffs are also uncertain. If publishers are risk-averse, they may choose to award their ad calls over a portfolio of advertising offers (see Bax, Chitrapura, Garg, and Gopalakrishnan \cite{bax009}). The publisher's role is similar to that of an investor in financial markets: just as investors allocate money to financial opportunities such as stocks and bonds, publishers allocate ad calls among potentially risky advertiser offers. 

Bax, Chitrapura, Garg, and Gopalakrishnan \cite{bax009} describe how to select a portfolio of ads on behalf of a risk-averse publisher, by applying classical mean-variance portfolio selection methods. (For background on these methods, refer to Markovitz \cite{markovitz52}, Lintner \cite{lintner65}, Sharpe \cite{sharpe64}, and Tobin \cite{tobin58}.) In auctions for online advertising, it is common to set the winner's expected payment to the expected value of the runner-up offer. (For more on second-price auctions, refer to Krishna \cite{krishna02} or Milgrom \cite{milgrom04}.) Under a portfolio allocation, there is no single winning offer and no single runner-up, so there is no clear notion of a second price. As a result, portfolio allocations require a different approach to pricing. As with second-price auctions, the portfolio pricing approach should have truth-telling as a dominant strategy. Also, intuitively, it should reward buyers who reduce risk for risk-averse sellers. 

The VCG (Vickrey-Clarke-Groves) mechanism is efficient, individually rational and incentive compatible. (For background on VCG mechanisms, refer to the original papers by Vickrey, Clarke, and Groves \cite{vickrey61,vickrey62,clarke71,groves73} or the books by Krishna \cite{krishna02} and Milgrom \cite{milgrom04}.) A generalized version of the VCG mechanism by Nisan \cite{nisan07} can accommodate side conditions and decision classes beyond auctions. This paper applies that framework to portfolio allocations, developing a mechanism that is efficient, incentive compatible, and rewards buyers who reduce seller risk.

This paper is organized as follows. Section \ref{sec:defns} reviews portfolio allocation and the generalized VCG mechanism framework. Section \ref{sec:results} shows how to apply that framework to portfolio allocations. Section \ref{sec:show} shows that the resulting mechanism has the desired properties and discusses a connection between the mechanism and second prices. Section \ref{sec:discuss} discusses potential directions for future work. 

\section{Background: Portfolio Allocation and Generalized VCG } \label{sec:defns}

This section reviews two concepts underlying a VCG mechanism for portfolio optimization. Subsection \ref{subsec:portfolio} reviews portfolio allocation. Subsection \ref{subsec:vcg} reviews VCG mechanisms.

\subsection{Portfolio Allocation}  \label{subsec:portfolio}

In classical portfolio methods, an investor has some funds to allocate among $n$ investment opportunities. Let random variables $X_{1},X_{2},.....,X_{n}$ be the returns for the investment opportunities. Let $\mbox{\boldmath$\mu$}=(\mu_{1},\mu_{2},.....,\mu_{n})$ be the expected returns: 

\be
\forall i:\mu_i = E \, X_i. 
\ee
Let $\mathbf{\Sigma}$ be the matrix of covariances of returns: 
\be
\forall i\neq j: \Sigma_{ij}= Cov(X_i, X_j),
\ee
and
\be
\forall i: \Sigma_{ii} = Var(X_i).
\ee

Let $\mathbf{w}=(w_{1},w_{2},.....,w_{n})$ be the weights in a portfolio allocation, with $w_i$ being the fraction of funds allocated to investment opportunity $i$. Then the expected return for the portfolio is 
\be
\mathbf{w}^T\mbox{\boldmath$\mu$} = \sum_{i=1}^{n} w_i\mu_i ,
\ee
and the variance of returns is 
\be
\mathbf{w}^T\mathbf{\Sigma}\mathbf{w}.
\ee

To compute a portfolio allocation, a profit-seeking but risk-averse investor may solve
\be
\mathbf{w}^{*} =  \underset{\mathbf{w}}{\operatorname{arg\,max}} (\mathbf{w}^T\mbox{\boldmath$\mu$} - q\mathbf{w}^T\mathbf{\Sigma}\mathbf{w}),
\ee
subject to
\be
\mathbf{w} \ge 0  \textnormal{ and } \mathbf{1}^T\mathbf{w} = 1. 
\ee
\noindent 
(For solution methods, refer to the original paper by Wolfe \cite{wolfe59}, the book by Franklin \cite{franklin80}, or another text on convex quadratic optimization techniques.) The investor selects a $q$ value to mediate a tradeoff between profit-seeking and risk-aversion. For example, $q=0$ expresses pure profit-seeking and no risk aversion.

\subsection{Generalized VCG Mechanism}  \label{subsec:vcg}
The VCG (Vickrey-Clarke-Groves) auction mechanism charges bidders based on the harm their participation causes other bidders. As a result, the VCG auction mechanism is a truthful mechanism, meaning that for each bidder, bidding their true value is a dominant strategy. In addition, the VCG mechanism is efficient, meaning that it maximizes the sum of utilities over participants. 

Because we work with portfolio allocations rather than auctions, in this paper we apply a generalized version of the VCG mechanism by Nisan \cite{nisan07}. In this setting, there is a set of outcomes $\mathcal{A}$ and a set of participants. Each participant $i$ submits valuations $v_i(a)$ for all outcomes $a\in \mathcal{A}$. The mechanism selects an outcome $a^*$ that maximizes the sum of valuations:
\be
a^* = \underset{a\in \mathcal{A}}{\operatorname{arg\,max}} \displaystyle\sum_{i}{v_i(a)}.
\ee
The mechanism charges each participant $i$: 
\be
p_i = h_i(v_{-i}) - \displaystyle\sum_{j\neq i}{v_j(a^*)}.
\ee
where $v_{-i}= (v_{1},.....,v_{i-1},v_{i+1},....,v_{n})$, and $h_{i}$ is a function that depends only on the valuations of the bidders other than bidder $i$. 

Like the original VCG mechanism, the generalized VCG mechanism is truthful and efficient. Informally, it is truthful because the charge to each participant does not depend on their stated valuation but rather on the harm caused to other participants. It is efficient because it selects the outcome that maximizes the sum of stated valuations, which are truthful as a dominant strategy for each participant.  (For more details, refer to Nisan \cite{nisan07}.)

\section{A VCG Mechanism for Portfolio Allocations}  \label{sec:results}

To apply the generalized VCG mechanism to portfolio allocation, consider the allocation as a process where:

\begin{enumerate}
\item The set of outcomes is
\be
\mathcal{A} = \{\mathbf{w}:\mathbf{w}^T \mathbf{1} = 1 ,  \mathbf{w} \geq \mathbf{0}\}. 
\ee  

\item There are $n$ advertising offers. Each offer is a participant submitting a valuation function
\be 
\forall i\in {1,.....,n}: v_i(\mathbf{w}) = w_i\mu_i.
\ee
In other words, the advertiser issuing offer $i$ values their allocation linearly in the fraction of the ad calls they receive.

\item There is an $(n+1)^{st}$ participant, representing the publisher's risk aversion. This participant submits the valuation function 
\be
v_{n+1}(\mathbf{w}) = -q\mathbf{w}^T\mathbf{\Sigma}\mathbf{w}.
\ee
\end{enumerate}

Then, when the general VCG mechanism selects an outcome to maximize the sum of valuations:
\be
a^* = \underset{a\in \mathcal{A}}{\operatorname{arg\,max}} \displaystyle\sum_{i}{v_i(a)},
\ee
it performs portfolio optimization:
\be
\mathbf{w}^* = \underset{\{\mathbf{w}:\mathbf{w}^T \mathbf{1}=1 , \mathbf{w}\geq \mathbf{0} \}}{\operatorname{arg\,max}}\mathbf{w}^{T}\mbox{\boldmath$\mu$} - q\mathbf{w}^T\mathbf{\Sigma} \mathbf{w},
\ee
where $\mathbf{w}^*$ is $a^*$ for the portfolio setting.

When the mechanism charges each participant
\be
p_i = h_i(v_{-i}) - \displaystyle\sum_{j\neq i}{v_j(a^*)},
\ee
for $ 1\leq i \leq n$, use
\be
h_i(v_{-i}) = [\displaystyle\max_{\{\mathbf{w}:\mathbf{w}^T \mathbf{1}=1, \mathbf{w}\geq \mathbf{0} ,w_{i}=0 \}} \mathbf{w}^T\mbox{\boldmath$\mu$} - q\mathbf{w}^T\mathbf{\Sigma}\mathbf{w}].
\ee
Hence charge each of the first $n$ participants:
\be
p_i = [\displaystyle\max_{\{\mathbf{w}:\mathbf{w}^T \mathbf{1}=1, \mathbf{w}\geq \mathbf{0} ,w_{i}=0 \}} \mathbf{w}^T\mbox{\boldmath$\mu$} -                          q\mathbf{w}^T\mathbf{\Sigma}\mathbf{w}]
\ee
\be
 - [\mathbf{w}^{*T}\mbox{\boldmath$\mu$}-                q\mathbf{w}^{*T}\mathbf{\Sigma}\mathbf{w}^{*} - w_i^{*}\mu_i],
\ee
which is the difference between the maximum sum of valuations over other participants when offer $i$ is removed from the system and the maximum sum of valuations over other participants when offer $i$ participates in the auction, submitting a valuation of $w_i\mu_i$. Publisher revenue is 

\be
\sum\limits_{i=1}^n{p_i}.
\ee

For the $(n+1)^{st}$ participant, who represents the publisher's risk aversion, let 
\be
h_{n+1}(v_{-(n+1)}) 
\ee
\be
= [\displaystyle\max_{\{\mathbf{w}:\mathbf{w}^T \mathbf{1}=1, \mathbf{w}\geq \mathbf{0} ,q=0 \}} \mathbf{w}^T\mbox{\boldmath$\mu$} - q\mathbf{w}^T\mathbf{\Sigma}\mathbf{w}]
\ee
\be
= \displaystyle\max_{\{\mathbf{w}:\mathbf{w}^T \mathbf{1}=1, \mathbf{w}\geq \mathbf{0}\}} \mathbf{w}^T\mbox{\boldmath$\mu$},
\ee
removing risk-aversion from the system. Then the mechanism ``charges" the $(n+1)^{st}$ participant:
\be
p_{n+1}= \displaystyle\max_{\{\mathbf{w}:\mathbf{w}^T \mathbf{1}=1 , \mathbf{w}\geq \mathbf{0} \}} \mathbf{w}^T\mbox{\boldmath$\mu$}-\mathbf{w}^{*T}\mbox{\boldmath$\mu$}.
\ee
The first term on the RHS is the expected revenue for a risk-neutral publisher. In this case, the allocation is to the highest bidder (or bidders if there is a tie). So:
\be
\quad p_{n+1}= \displaystyle\max_{i\in \{1,.....,n \}}\mu_i - \mathbf{w}^{*T}\mbox{\boldmath$\mu$},
\ee
which is the expected revenue foregone by the publisher due to risk aversion. 

Note that prices $p_1,.....,p_n$ are per-allocation prices. To convert to charges per ad call, divide each offer's price by the number of ad calls in the offer's allocation. This per-ad call charge applies directly to offers that pay on an ad call basis. For offers that pay based on user responses, divide the per-ad call charge by the estimated response rate to compute a per-response charge.

This section applied VCG pricing to portfolio allocations without uncertainty and using standard portfolio allocation notation. Appendix A applies VCG pricing with uncertainty, following the notation in \cite{bax009}.

\section{Properties and Connection to Second Prices}
\label{sec:show}
The mechanism developed in Section \ref{sec:results} is efficient if bidders report their true valuations, since the mechanism maximizes the sum of reported valuations. This section begins with a proof that the mechanism is truthful. Next we show that it is individually rational, meaning that each buyer receives nonnegative value from participation. Then we outline a connection between the mechanism's prices and second prices. 

\subsection{Truthfulness}
This subsection shows that bidding one's own value is the optimal strategy for each of the $n$ bidders. The utility derived from participation in the auction for bidder $i$, denoted $u_i$, is the difference between the bidder's value for the allocation received and the bidder's payment. In other words, 

\be
u_i(\mathbf{w}) = w_i\mu_i - p_i.
\ee

If all bidders their true valuations $\mbox{\boldmath$\mu$}$, then the allocation is $\mathbf{w}^*$. So the utility for bidder $i$ is

\be
u_i(\mathbf{w}^*) =  w_i^*\mu_i - [h_i(v_{-i}) - (\mathbf{w}^{*T}\mbox{\boldmath$\mu$}-q\mathbf{w}^{*T}\mathbf{\Sigma}\mathbf{w}^*-w^*_i\mu_i)] 
\ee
\be
= -h_i(v_{-i}) + \mathbf{w}^{*T}\mbox{\boldmath$\mu$}-q\mathbf{w}^{*T}\mathbf{\Sigma}\mathbf{w}^*.
\ee

Now suppose bidder $i$ decides to bid $v_i = \mu_i + \Delta$ instead of $\mu_i$, while all other bidders bid their true valuations. Then the portfolio allocation uses $\mbox{\boldmath$\mu$}(\Delta) = (\mu_1, ..., \mu_{i-1}, \mu_i + \Delta, \mu_{i+1}, ..., \mu_n)$, to produce an allocation $\mathbf{w}^\Delta$. So
\be
u_i(\mathbf{w}^\Delta) 
\ee
\be
= w^\Delta_i\mu_i - [ h_i(v_{-i}) - (\mathbf{w}^{\Delta T}\mbox{\boldmath$\mu$}(\Delta)-q\mathbf{w}^{\Delta T}\mathbf{\Sigma}\mathbf{w}^\Delta-w^\Delta_i(\mu_i + \Delta))]
\ee
\be
= w^\Delta_i\mu_i - [ h_i(v_{-i}) - (\mathbf{w}^{\Delta T}\mbox{\boldmath$\mu$} + w^\Delta_i \Delta -q\mathbf{w}^{\Delta T}\mathbf{\Sigma}\mathbf{w}^\Delta-w^\Delta_i(\mu_i + \Delta))]
\ee
\be
=  -  h_i(v_{-i}) + \mathbf{w}^{\Delta T}\mbox{\boldmath$\mu$}  -q\mathbf{w}^{\Delta T}\mathbf{\Sigma}\mathbf{w}^\Delta
\ee

Note that $h_i(v_{-i})$ is independent of $v_i$, so it cancels out in the difference between $u_i(\mathbf{w}^*)$ and $u_i(\mathbf{w}^\Delta)$, leaving:
\be
u_i(\mathbf{w}^*) - u_i(\mathbf{w}^\Delta) 
\ee
\be
= [\mathbf{w}^{*T}\mbox{\boldmath$\mu$}-q\mathbf{w}^{*T}\mathbf{\Sigma}\mathbf{w}^*] - [\mathbf{w}^{\Delta T}\mbox{\boldmath$\mu$}-q\mathbf{w}^{\Delta T}\mathbf{\Sigma}\mathbf{w}^\Delta] \ge 0,
\ee
the last inequality following by definition of $\mathbf{w}^*$.

\subsection{Individual Rationality}
Now consider whether the mechanism is individually rational for the $n$ bidders. Since $\mathbf{w}^{*}$ is chosen from the set $\Gamma = {\{\mathbf{w}:\mathbf{w}^T \mathbf{1}=1, \mathbf{w}\geq \mathbf{0}\}}$, we must have the maximized value of the objective at least as large as when $\mathbf{w}$ is chosen from the set $\Gamma$ with the additional restriction that $w_{i}=0$. So:
\be
\mathbf{w}^{*T}\mbox{\boldmath$\mu$}-                q\mathbf{w}^{*T}\mathbf{\Sigma}\mathbf{w}^{*} \geq [\displaystyle\max_{\{\mathbf{w}:\mathbf{w}^T \mathbf{1}=1, \mathbf{w}\geq \mathbf{0} ,w_{i}=0 \}} \mathbf{w}^T\mbox{\boldmath$\mu$} -                          q\mathbf{w}^T\mathbf{\Sigma}\mathbf{w}].
\ee
Thus, the expected payoff for advertiser $i$ from participating in the auction is nonnegative.

\subsection{Convergence to Second Prices}
To see a connection with a second-price mechanism, consider what happens when $q=0$, indicating that the publisher is risk-neutral. In this case, $w_{i}^{*}=1$ for the offer $i$ having the greatest $\mu_{i}$. (For simplicity, assume the maximum is unique). So all ad calls are awarded to the offer having the highest expected revenue. The price for the winning offer $i$ is:
\be
\displaystyle\max_{\{\mathbf{w}:\mathbf{w}^T \mathbf{1}=1 , \mathbf{w}\geq \mathbf{0},w_{i}=0 \}} \mathbf{w}^T\mbox{\boldmath$\mu$},
\ee
which is simply the expected revenue from awarding all ad calls to the second-highest bidder. The price for the other offers is zero, and the price for the publisher's (lack of) risk aversion is zero.

\section{Discussion}
\label{sec:discuss}
This paper describes a VCG mechanism to set prices for online advertisers under portfolio allocations for risk-averse publishers. The basic idea is to introduce an extra participant that represents the publisher's risk aversion. Then the generalized VCG outcome selection method matches the portfolio allocation method, and the VCG pricing method reduces prices for advertisers who reduce publisher risk.

The basic idea can be generalized to introduce ``social costs'' or ``future costs'' into the VCG mechanism. For example, when a government allocates bandwidth among telecommunications companies, the government's goals may include raising revenue and limiting how much bandwidth is awarded to any single company, in order to maintain competition. The market-maker could introduce an extra participant, with valuations based on how much competition remains after allocations. Applying the general VCG mechanism could maximize an objective function that balances revenue for the government with competition among bandwidth owners after the allocation, reducing prices for bandwidth buyers who maintain competition.

For online advertising, one direction for future work is to introduce valuations for factors such as the value of learning response rates and the value of ad impact on user experience. The value of learning is based on the probability that showing an ad will lead to learning that it has a high enough response rate to warrant allocating more ad calls to it in the future, increasing future publisher revenue, as discussed by Li, Mahdian, and McAfee \cite{li10}. (Learning also reduces uncertainty about response rate, decreasing future risk to publisher revenue.) Ad impact on user experience is important because a publisher who shows high-revenue, but disturbing, ads may drive away users, reducing future ad calls and revenue, as explained by Abrams and Schwarz \cite{schwarz08}. Likewise, ads that users appreciate may spur users to visit websites more often and to recommend them to friends. For both the value of learning and the value of ad impact on user experience, the expected net present value and variance may be estimated, but with some uncertainty. So these factors may be included in the revenue and variance inputs to the portfolio allocation model.

Several estimation correction methods have been proposed for portfolio allocation in financial markets. It would be interesting to explore how using these methods for online advertising markets would affect allocations and VCG prices. These methods include James-Stein and other shrinkage methods (see Bock \cite{bock75}, Brown \cite{brown66}, Stein \cite{stein55}, and James and Stein \cite{james61}), empirical Bayes estimation (see Jorion \cite{jorion86}), other methods to correct for uncertainty (see Jobson, Korkie, and Ratti \cite{jobson79} and Vasicek \cite{vasicek73}), and robust optimization methods (see Scherer \cite{scherer07a,scherer07b}, and Tutuncu \cite{tutuncu04}).

Another direction for future work is to accomodate more complex bidding behavior, including allowing advertisers to set budgets or, more generally, to express varying marginal values for fractions of the allocation. If advertisers express budget constraints in terms of a limit on the fraction of ad calls they will accept, then these constraints translate directly into linear constraints in the quadratic program to solve for the portfolio allocation. Also, the VCG pricing mechanism remains the same, with the limits on fractions of allocations included in the specification of the feasible set of outcomes $\mathcal{A}$. 

However, if the advertisers express budgets in terms of limits on money to be spent, a few complications may arise. One complication is that the VCG prices depend on the allocations, so the market-maker cannot simply convert a spending limit for an advertiser into a limit on the fraction of ad calls allocated to the advertiser. So an iterative procedure may be required. Another complication is that if advertisers pay per response, then the number of responses that will be realized per ad call is not known when the ad calls are allocated.

\section*{Acknowledgement}
We thank Michael Ostrovsky for his valuable advice and feedback.

\appendix
\section{Applying VCG Pricing to QMAP}
We will apply the generalized VCG mechanism to the QMAP problem from Bax, Chitrapura, Garg, and Gopalakrishnan \cite{bax009}:
\be
\displaystyle\min_{\mathbf{k}}\mathbf{k}^T\mathbf{A}\mathbf{k} + \mathbf{b}^T\mathbf{k}- q\mathbf{c}^T\mathbf{k}.
\ee
where $\mathbf{A}$ and $\mathbf{b}$ express uncertainty and randomness, and $\mathbf{c}$ expresses expected returns. Please refer to \cite{bax009} for details.

Transform to an equivalent problem by substituting $1/q$ for $q$, multiplying by $q$, and reversing the signs:
\be
\displaystyle\max_{\mathbf{k}} \mathbf{c}^T\mathbf{k} - q(\mathbf{k}^T\mathbf{A}\mathbf{k} + \mathbf{b}^T\mathbf{k}).
\ee
Then 
\be
\mathcal{A} = \{\mathbf{k}:\mathbf{k}^T \mathbf{1} = m ,  \mathbf{k} \geq \mathbf{0}\},
\ee
where $m$ is the number of available ad calls (The allocations $\mathbf{k}$ are in terms of number of ad calls, rather than the fraction of ad calls in allocations $\mathbf{w}$). Also:
\be 
\forall i\in {1,.....,n}: v_i(\mathbf{k}) = c_{i}k_{i},
\ee
since $c_{i}$ is the expected revenue per ad call. For the risk-aversion participant,
\be v_{n+1}(\mathbf{k}) = -q(\mathbf{k}^T\mathbf{A}\mathbf{k} + \mathbf{b}^T\mathbf{k}).
\ee
So the allocation is 
\be
\mathbf{k}^* = \underset{\{\mathbf{k}:\mathbf{k}^T \mathbf{1}=m , \mathbf{k}\geq \mathbf{0} \}}{\operatorname{arg\,max}}\mathbf{c}^T\mathbf{k} - q(\mathbf{k}^T\mathbf{A}\mathbf{k} + \mathbf{b}^T\mathbf{k}),
\ee
and the offer prices are
\be
\forall i\in {1,.....,n}:
\ee
\be
p_i = [\displaystyle\max_{\{\mathbf{k}:\mathbf{k}^T \mathbf{1}=m, \mathbf{k}\geq \mathbf{0}, k_i = 0 \}} \mathbf{c}^T\mathbf{k} - q(\mathbf{k}^T\mathbf{A}\mathbf{k} + \mathbf{b}^T\mathbf{k})] 
\ee
\be
- [\mathbf{c}^T\mathbf{k}^{*} - q(\mathbf{k}^{*T}\mathbf{A}\mathbf{k}^{*} + \mathbf{b}^T\mathbf{k}^{*})].
\ee

\bibliographystyle{abbrv}
\bibliography{bib}
\end{document}